\documentclass[11pt,a4paper,epsf,epsfig,psfrag]{article}
\usepackage{jheppub}
\usepackage{amsmath,graphicx,amsfonts, mathrsfs,amssymb}
\usepackage{amsmath,epsfig}
\usepackage{amssymb,amsfonts}
\usepackage{latexsym}
\usepackage{epsfig}
\usepackage{pdfsync}
\newbox\pippobox

\def\be{\begin{equation}}
\def\ee{\end{equation}}
\def\bea{\begin{eqnarray}}
\def\eea{\end{eqnarray}}

\def\lag{\langle}
\def\rag{\rangle}

\newcommand{\beq}{\begin{equation}}
\newcommand{\eeq}{\end{equation}}
\newcommand{\beqa}{\begin{eqnarray}}
\newcommand{\eeqa}{\end{eqnarray}}
\newcommand{\beqar}{\begin{eqnarray*}}
\newcommand{\eeqar}{\end{eqnarray*}}

\renewcommand{\eqref}[1]{(\ref{#1})}

\catcode`\@=12

\def\lag{\langle}
\def\rag{\rangle}

\title{Temperature dependent transport coefficients in a dynamical holographic QCD model}
\author[a,b]{Danning Li,}
\author[c,a]{Song He,}
\author[b,d]{Mei Huang}

\affiliation[a]{State Key Laboratory of Theoretical Physics,
Institute of Theoretical Physics, Chinese Academy of Science,
Beijing 100190, P. R. China } \affiliation[b]{Institute of High
Energy Physics, Chinese Academy of Sciences, Beijing 100049, P.R. China}
\affiliation[c]{Yukawa Institute for Theoretical Physics, Kyoto University, Kitashirakawa Oiwakecho,
Sakyo-ku, Kyoto 606-8502, Japan}
\affiliation[d]{Theoretical Physics Center for Science
Facilities, Chinese Academy of Sciences, Beijing 100049, P.R. China}
\emailAdd{lidn@itp.ac.cn}\emailAdd{hesong17@gmail.com}\emailAdd{huangm@ihep.ac.cn}

\abstract{We investigate temperature dependent behavior of various transport coefficients in a dynamical holographical QCD model. We show the nontrivial  temperature dependent behavior of the transport coefficients, like bulk viscosity, electric conductivity as well as jet quenching parameter, and it is found that all these quantities reveal information of the phase transition. Furthermore, with introducing higher derivative corrections in 5D gravity, the shear viscosity over entropy density ratio also shows a valley around phase transition, and it is found that the shear viscosity over entropy density ratio times the jet quenching over temperature cubic ratio almost remains as a constant above phase transition, and the value is two times larger than the perturbative result in Phys.Rev.Lett.99.192301(2007). }


\begin{document}
\maketitle
\section{Introduction}

Quantum chromodynamics (QCD) is the fundamental theory of the strong interactions. The study of the QCD phase
transitions and phase structure at extreme conditions are always the most important topics of high energy nuclear physics.
Ultrarelativistic heavy ion collisions (HIC) provide a unique controllable experimental tool to investigate properties of
nuclear matter at high temperature and small baryon density. It is believed that at the Relativistic Heavy Ion collider (RHIC)
and the Large Hadron Collider (LHC), the quarks and gluons can become deconfined and form hot quark-gluon plasma (QGP)
\cite{Shuryak:1980tp,McLerran:1986zb}.  The main target of HIC is to study the equation of state (EoS) and transport
properties of hot/dense QCD matter.

The pressure gradient of the created fire ball in HIC would drive it to expand and cool down. Finally, the whole system would
hadronize and only measurable mesons and baryons are left to the detectors.  The real time evolution of the plasma could be
studied in effective models, such as Boltzmann transport, hydrodynamics and hadronic cascade \cite{Molnar:2001ux,Teaney:2000cw,Bleicher:1999xi,Bass:1998ca,Heinz:2013th,Gale:2013da}.
In these effective models, transport coefficients, such as bulk/shear viscosity, jet quenching parameter
and electric conductivity, appear as parameters encoding the long wavelengths and low frequency fluctuation
dynamics of the underlying quantum field theory and are used to characterize the non-equilibrium dynamics
of the system.

Shear viscosity $\eta$ characterizes the ability of a system to recover equilibrium after shear mode perturbation.
In microscopic level, the value of shear viscosity over entropy density ratio $\eta/s$ is related to the interaction
strength of interparticles in a system. In general, a stronger interaction strength corresponds to a smaller $\eta/s$
ratio. At weakly coupling, the perturbative QCD calculation gives the result of $\eta\propto1/\alpha_s^2 \ln \alpha_s$
\cite{Arnold:2000dr}, where $\alpha_s$ is the strong coupling constant. At strong coupling, Lattice QCD simulation
showed that $\eta/s$ for the purely gluonic plasma is rather small and in the range of $0.1-0.2$ \cite{LAT-etas}.
The studies from AdS/CFT correspondence \cite{Maldacena:1997re,Gubser:1998bc,Witten:1998qj} give the lower
bound of $\eta/s= \frac{1}{4\pi}$ \cite{Kovtun:2004de}, which is very close to the value used to fit the RHIC data
of elliptic flow $v_2$ \cite{Song:2008hj,Hydro, Hydro-Teaney}. Therefore, it is now widely believed that the system
created at RHIC and LHC behaves as a nearly "perfect" fluid and is strongly coupled \cite{RHIC-EXP,RHIC-THEO}.

Like shear viscosity, bulk viscosity $\zeta$ describes how fast a system returns to equilibrium under a uniform expansion.
In the perturbative region, the bulk viscosity $\zeta$ is very small and the leading dependence on $\alpha_s$ is
$\zeta\propto \alpha_s^2/\ln \alpha_s ^{-1}$ \cite{Arnold:2006fz}. However, $\zeta/s$ is shown to rise sharply near $T_c$
in studies from different approaches, for example, Lattice QCD \cite{LAT-xis-KT,LAT-xis-Meyer,correlation-Karsch}, the
linear sigma model \cite{bulk-Paech-Pratt}, the Polyakov-loop linear sigma model \cite{Mao:2009aq} and the real scalar model \cite{Li-Huang}. The near phase transition enhancement of bulk viscosity corresponds to the peak of trace anomaly around $T_c$,
which shows the equation of state is highly non-conformal \cite{LAT-EOS-G, LAT-EOS-Nf2} around phase transition.

Phenomenologically, shear/bulk viscosities are used to study the collective flow behavior of the hot/dense nuclear matter created in HIC. Besides, the suppression of large transverse momentum hadrons emission in central collisions also attracts many attentions \cite{jet-quenching}. This suppression behavior is normally referred to as jet quenching, which characterizes the squared average transverse momentum exchange between the medium and the fast parton per unit path length~\cite{Baier:1996kr}. Current understanding on jet quenching is that it is caused by gluon radiation induced by multiple collisions of the leading parton with color charges in the near-thermal medium \cite{Baier:1996kr,energy-loss,GuoW}. Therefore, it is possible to extract the properties of the created hot dense matter from jet quenching of energetic partons passing through the medium.

Besides the above three transport coefficients, other transport coefficients are also attracting more and more attentions. For example,
due to the strong electric field created in the collision region, electric conductivity is now becoming more and more interesting \cite{Cassing:2013iz,Marty:2013ita}. The electric conductivity describes the response of the system to external electric field and is
responsible for the production of the collective flow of charged particles \cite{Tuchin:2013ie}.

Except for the phenomenological application of transport coefficients, another interesting issue is to study the temperature dependent
transport coefficients and investigate whether transport quantities can characterize phase transition.  As mentioned above, the near $T_c$ enhancement of bulk viscosity $\zeta$ is due to the rapid change of degrees of freedom from phase transition. It has been proposed
by the authors of  \cite{Lacey:2006bc} to use the shear viscosity over entropy density ratio as a signal of the critical end point(CEP).
It has also been shown that $\eta /s$ is suppressed near the critical temperature in the semi  quark gluon plasma \cite{Hidaka:2009ma},
and a cusp, a jump at $T_c$ and a shallow valley around $T_c$ in $\eta /s$ can characterize first-, second-order phase transitions
and crossover \cite{etas-scalar}. If the perturbative calculation based conclusion in \cite{Majumder:2007zh} that small $\eta/s$ implies
large jet quenching parameter can be extended to strong coupled region, then we can expect that the temperature dependence behavior of jet quenching parameter can also contain the information of phase transition. Near the phase transition point, the nonperturbative dynamics dominates and the perturbative calculation becomes invalid. It's quite convenient to use the holographic method based on the AdS/CFT correspondence to study the dynamical properties, while Lattice calculation in this area still suffer large uncertainty. Within holographic framework, our previous work \cite{Li:2014hja} do show that near the transition point there would be nontrivial enhancement of $\hat{q}/T^3$, which is closely related to the phase transition.

In this work, we will investigate the temperature dependent behavior of shear viscosity, bulk viscosity and electric conductivity in the
framework of dynamical holographic QCD model \cite{Li:2014hja}.It has been proved that in any isotropic Einstein gravity background the shear viscosity over entropy density ratio is always $\frac{1}{4\pi}$ \cite{Policastro:2001yc,Buchel:2003tz}. Only with the higher derivative corrections\cite{Kats:2007mq,Brigante:2007nu,Brigante:2008gz} or in a anisotropic gravity background\cite{Natsuume:2010ky,Erdmenger:2010xm} one can get deviation from $\frac{1}{4\pi}$. In order to introduce temperature dependence behavior of $\eta/s$, we will follow the formalism in \cite{Myers:2009ij,Cremonini:2011ej,Cremonini:2012ny} and continue our studies in \cite{Li:2014hja}. The paper would mainly focus on the nontrivial near $T_c$ behavior of the transport coefficients.
The paper are organized as followings. In Sec.\ref{sec-dhqcd} we give a brief review on the dynamical holographic QCD model which works quite well, both at zero temperature for meson spectra \cite{Li:2012ay,Li:2013oda} and at finite temperature for the equation of state and jet quenching properties \cite{Li:2014hja}.  In Sec.\ref{sec-tans-coeff}, we will show the results of temperature dependent transport coefficients in our dynamical holographic QCD model. Finally, a short summary would be given in Sec.\ref{sec-sum}.

\section{Dyanmical holographic QCD model}
\label{sec-dhqcd}

Quantum chromodynamics (QCD), with quarks and gluons as its basic degrees of freedom, is widely accepted as the fundamental theory of the strong interaction. In short distance or ultraviolet (UV) regime, the QCD coupling constant $\alpha_s$ is small and results from the QCD perturbative calculations agree well with the experimental data. Unfortunately, in large distance or infrared (IR) regime, the coupling constant evolves to a large value and perturbative studies on QCD vacuum as well as hadron properties and other nonperturbative processes become invalid. Several non-perturbative methods have been developed, in particular lattice QCD, Dyson-Schwinger equations (DSEs), and functional renormalization group equations (FRGs), which are based on first principle derivation. Recently, the conjecture of the gravity/gauge duality \cite{Maldacena:1997re,Gubser:1998bc,Witten:1998qj} provide a new hopeful method to tackle with the strong coupling problem of strong interaction.

The weak version of the conjecture is the duality between a quantum field theory (QFT) in d-dimensions and a quantum
gravity in (d + 1)-dimensions. More importantly, when the QFT is strongly-coupled theory, the gravitational description becomes classical and usually a saddle point approximation in gravity side is sufficient to reach the expected accuracy. The extra dimension in the gravity side can play the role of the energy scale of the renormalization group (RG) flow in the QFT \cite{Adams:2012th}.

\begin{figure}[h]
\begin{center}
\epsfxsize=7.5 cm \epsfysize=5.5 cm \epsfbox{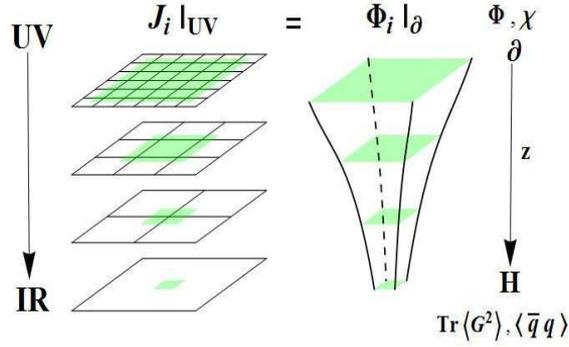}
\end{center}
\caption{Schematic plots of the correspondence between $d$-dimension QFT and $d+1$-dimension gravity as shown
in \cite{Adams:2012th} (Left-handed side). Dynamical holographic QCD model resembles
RG from UV to IR (Right-handed side): at UV boundary the dilaton bulk field $\Phi(z)$
and scalar field $X(z)$ are dual to the dimension-4 gluon operator $Tr\langle G^2\rangle$ and dimension-3
quark-antiquark operator $\langle \bar{q}q\rangle$, which develop condensates at infrared regime.  }
 \label{fig:RGflow}
\end{figure}

A dynamical holographic QCD model \cite{Li:2012ay,Li:2013oda} has been constructed by potential reconstruction approach \cite{Li:2011hp}. In this model, authors introduce the dilaton field $\Phi(z)$ to describe the gluon dynamics and the scalar field $X(z)$ to mimic chiral dynamics. The model describes the glueball and the light meson spectral quite well. Evolution of these fields in 5D resemble the renormalization group from ultraviolet (UV) to infrared (IR) as shown in Fig.\ref{fig:RGflow} \cite{Li:2013xpa}.

For pure gluon system, to break the conformal symmetry of the pure AdS background to mimic realistic QCD, we introduce the dilaton field $\Phi(z)$ and consider the gluon dynamics in the graviton-dilaton coupled system, which is regarded as a quenched dynamical holographic QCD model. The 5D action for the graviton-dilaton system in string frame reads
\begin{eqnarray}
 S_G=\frac{1}{16\pi G_5}\int
 d^5x\sqrt{g_s}e^{-2\Phi}\left(R_s+4\partial_M\Phi\partial^M\Phi-V^s_G(\Phi)\right).
\label{action-gluon}
\end{eqnarray}
Here $G_5$ is the 5D Newton constant, $g_s$ is the 5D metric determinant and $\Phi$ and $V_G^s$ are the dilaton field and dilaton potential in the string frame respectively. Under the frame transformation
\begin{equation}
g^E_{mn}=g^s_{mn}e^{-2\Phi/3}, ~~ V^E_G=e^{4\Phi/3}V_{G}^s,
\end{equation}
the Einstein frame action of Eq.(\ref{action-gluon}) takes the form
\begin{eqnarray}\label{graviton-dilaton-E}
S_G^E=\frac{1}{16\pi G_5}\int d^5x\sqrt{g_E}\left(R_E-\frac{4}{3}\partial_m\Phi\partial^m\Phi-V_G^E(\Phi)\right).
\end{eqnarray}

At zero temperature, the metric ansatz is often chosen to be
\begin{eqnarray}\label{metric-ansatz}
ds^2=e^{2A_s(z)}(dz^2+\eta_{\mu\nu}dx^\mu dx^\nu).
\end{eqnarray}
In this paper, the capital letters like $M,N$ stand for all the 5D coordinates $(0,1,..,4)$, and the Greek indexes stand for the 4D coordinates $(0,...,3)$ only. We would use the sign convention $\eta^{00}=\eta_{00}=-1,\eta^{ij}=\eta_{ij}=\delta_{ij}$ in the metric.

Starting with the metric ansatz Eq.(\ref{metric-ansatz}), the equations of motion can be derived as
\begin{eqnarray}
-A_s^{''}-\frac{4}{3}\Phi^{'}A_s^{'}+A_s^2+\frac{2}{3}\Phi^{''}=0, \\
\Phi^{''}+(3A_s^{'}-2\Phi^{'})\Phi^{'}-\frac{3}{8}e^{2A_s-\frac{4}{3}\Phi}\partial_\Phi V_G^E(\Phi)=0.
\end{eqnarray}

Generally speaking, there are three different approaches \cite{Li:2013oda} to solve this graviton-dilaton coupled system in the literature,
\begin{enumerate}
\item  Input a certain configuration of the field(s), for example $\Phi(z)$, to solve the metric structure
and the potential(s) of the field(s) \cite{Csaki:2006ji,Li:2012ay,Li:2013oda},
\item Input the potential(s) of the field(s) to solve the metric and the field(s) \cite{Gubser:2008ny,GKN},
\item Input the metric structure to solve the field(s) and the potential(s) of the field(s) \cite{Li:2011hp}.
\end{enumerate}

In this paper, we will follow the first approach and take the dilaton field configuration as an input. Firstly, we should choose profile of dilaton field to construct dynamical model.
The dilaton field is dual to the dimension-4 gauge invariant gluon condensate ${\rm Tr}\lag G^2 \rag $ and UV behavior of dilaton is
\begin{equation}
\Phi(z)\overset{z\rightarrow0}{\rightarrow} \mu_{G^2}^4 z^4,
\end{equation}
while IR behavior of dilaton is
\begin{equation}
\Phi(z)\overset{z\rightarrow\infty}{\rightarrow} \mu_G^2 z^2.
\end{equation} IR behavior of dilaton is necessary for the linear confinement.

In \cite{Li:2013oda,Li:2014hja}, we interpolate these two asymptotic behaviors by taking the following dilaton profile
\begin{equation}
\Phi(z)=\mu_G^2z^2\tanh(\mu_{G^2}^4z^2/\mu_G^2).
\label{mixed-dilaton}
\end{equation}
When the parameter $\mu_{G^2}$ goes to infinity, the model reduces to the quadratic form
\begin{equation}
\Phi(z)=\mu_G^2z^2,
\label{quadratic-dilaton}
\end{equation}
which can be regarded as a dynamical extension of the KKSS model \cite{Karch:2006pv}. The previous studies also showed that when $\mu_G$ is very large, there are no much differences between models with dilaton profile Eq.(\ref{mixed-dilaton}) and with dilaton profile Eq.(\ref{quadratic-dilaton}). The main reason is that meson spectral and equations of state are much more sensitive to the IR behavior. Since we will focus on the temperature dependence transport coefficients and for simplicity, we would only consider the quadratic dilaton profile Eq.(\ref{quadratic-dilaton}) in this work.

By self-consistently solving the above equations, the metric $A_s$ will be deformed in IR region by the background dilaton field. The scalar glueball spectra agree with lattice data quite well in the quenched dynamical model \cite{Li:2013oda}.

Secondly, we would like to mimic the flavor dynamics with one additional scalar. The total 5D action becomes the following graviton-dilaton-scalar system,
\begin{eqnarray}
 S=S_G + \frac{N_f}{N_c} S_{KKSS},
\end{eqnarray}
with $S_G$ the 5D action with form of Eq.(\ref{action-gluon}). $S_{\text{KKSS}}$ the 5D action
for mesons takes form as the KKSS model \cite{Karch:2006pv}.
\begin{eqnarray}
S_{KKSS}&=&-\int d^5x
 \sqrt{g_s}e^{-\Phi}Tr(|DX|^2+V_X(X^+X, \Phi) \nonumber \\
 && ~~+\frac{1}{4g_5^2}(F_L^2+F_R^2)).
\end{eqnarray}
Under metric ansatz as Eq.(\ref{metric-ansatz}), the equations of motion for such a system is also easily derived,
\begin{eqnarray}
 -A_s^{''}+A_s^{'2}+\frac{2}{3}\Phi^{''}-\frac{4}{3}A_s^{'}\Phi^{'}
 -\frac{\lambda_0}{6}e^{\Phi}\chi^{'2}&=&0, \label{Eq-As-Phi} \\
 \Phi^{''}+(3A_s^{'}-2\Phi^{'})\Phi^{'}-\frac{3\lambda_0}{16}e^{\Phi}\chi^{'2} & &
 \nonumber \\
 -\frac{3}{8}e^{2A_s-\frac{4}{3}\Phi}\partial_{\Phi}\left(V_G(\Phi)
 +\lambda_0 e^{\frac{7}{3}\Phi}V_C(\chi,\Phi)\right)&=&0, \label{Eq-VG}\\
 \chi^{''}+(3A_s^{'}-\Phi^{'})\chi^{'}-e^{2A_s}V_{C,\chi}(\chi,\Phi)&=&0. \label{Eq-Vc}
\end{eqnarray}
Here we have used the redefinition $V_C=  Tr(V_X)$ and $V_{C,\chi}=\frac{\partial V_C}{\partial \chi}$,$ \frac{16\pi G_5 N_f}{L^3 N_c}\rightarrow \lambda_0 $.

Finally, graviton-dilaton-scalar system has been set up already with considering both the chiral condensate and gluon dynamics in the vacuum. Here we also consider the mixing effects between the chiral condensate and gluon condensate is also important to produce the correct light flavor meson spectra \cite{Li:2013oda}. In the following series sections, we would like to study some properties of this model.

\subsection{Black hole solution and equation of state}
\label{sec-phase-eos}

In previous section, we have reviewed how to construct the dynamic model for zero temperature.
In this section we will first briefly review the thermodynamics in our dynamical holographic QCD model (For details, please refer to Refs.\cite{Li:2014hja}). We will focus on the confinement/deconfinement phase transition in the graviton-dilaton system defined in Eq.(\ref{action-gluon}). Chiral phase transition is also interesting and left to the future work.

To study the thermodynamics in 4D field theory from AdS/CFT, it's natural to introduce black hole in the 5D gravity side firstly. The finite temperature metric ansatz in string frame becomes
\begin{equation} \label{metric-stringframe}
ds_S^2=
e^{2A_s}\left(-f(z)dt^2+\frac{dz^2}{f(z)}+dx^{i}dx^{i}\right).
\end{equation}

Under this metric ansatz, from the Einstein equations of $(t,t), (z,z)$ and $(x_1, x_1)$ components and the dilaton field equation we get the following equations of motion,
\begin{eqnarray}
 -A_s^{''}+A_s^{'2}+\frac{2}{3}\Phi^{''}-\frac{4}{3}A_s^{'}\Phi^{'}&=&0, \label{Eq-As-Phi-T} \\
 f''(z)+\left(3 A_s'(z) -2 \Phi '(z)\right)f'(z)&=&0,\label{Eq-As-f-T}\\
 \frac{8}{3} \partial_z
\left(e^{3A_s(z)-2\Phi} f(z)
\partial_z \Phi\right)-
e^{5A_s(z)-\frac{10}{3}\Phi}\partial_\Phi V_G^E&=&0,
\end{eqnarray}
with $V^E_G=e^{4\Phi/3}V_{G}^s$.

In terms of AdS/CFT, we should impose the asymptotic AdS boundary condition $A_s\overset{z\rightarrow0}{\rightarrow}0, \text{ } \text{ }f\overset{z\rightarrow0}{\rightarrow}1$. We also
require $\Phi$ to be finite at $z=0, z_h$. Where the black-hole horizon $z_h$ is determined by $f(z_h)=0$. Fortunately, the solution of the black-hole background takes the following semi-analytical form of
\begin{eqnarray} \label{solu-f}
f(z)= 1- f_{c}^h \int_0^{z} e^{-3A_s(z^{\prime})+2\Phi(z^{\prime})} dz^{\prime},
\end{eqnarray}
with
\begin{eqnarray}\label{fc}
f_{c}^h= \frac{1}{\int_0^{z_h} e^{-3A_s(z^{\prime})+2\Phi(z^{\prime})} dz^{\prime} }.
\end{eqnarray}
The temperature of the solution would be identified with the Hawking temperature
\begin{equation} \label{temp}
T =\frac{e^{-3A_s(z_h)+2\Phi(z_h)}}{4\pi \int_0^{z_h} e^{-3A_s(z^{\prime})+2\Phi(z^{\prime})} dz^{\prime} }.
\end{equation}

By taking the dilaton profile of Eq.(\ref{quadratic-dilaton}), the metric prefactor $A_s$ could be solved analytically as
\begin{equation}\label{As-sol}
A_s(z) =\log(\frac{L}{z})-\log(_0F_1(5/4,\frac{\mu_G^4z^4}{9}))+\frac{2}{3}\mu_G^2z^2,
\end{equation}
where $L$ is the AdS radius. Since in the following calculation, only the combination of $G_5/L$ is relative, we will set $L$ to be $1$ in this paper. $\mu_G$ is a free parameter of the model, and to fix the phase transition temperature at around $255{\rm MeV}$, $\mu_G$ will be fix to $0.75 {\rm GeV}$ later.

From Eq.(\ref{temp}) and Eq.(\ref{As-sol}), we can get the relation between temperature $T$ and the black hole horizon $z_h$. Taking $\mu_G=0.75 {\rm GeV}$, we can get the critical temperature for deconfinement phase transition is $T_c=255 {\rm MeV}$.
The black hole entropy density $s$ could be easily read from the Bekenstein-Hawking formula,
\begin{eqnarray}
s=\frac{1}{4G_5}e^{3A_s(z_h)-2\Phi(z_h)}.
\end{eqnarray}

\begin{figure}[h]
\begin{center}
\epsfxsize=6.5 cm \epsfysize=6.5 cm \epsfbox{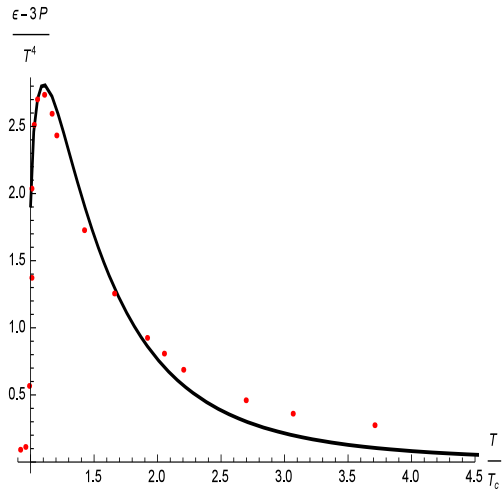}
\end{center}
\caption[]{The trace anomaly $\epsilon-3p$ as a function of $T_c$ scaled temperature $T/T_c$ for $G_5=1.25$ and $\mu_G=0.75 {\rm GeV}$(Solid black line). The red dots are the pure SU(3) lattice data from \cite{LAT-EOS-G}.} \label{TraceA}
\end{figure}

The pressure density $p$, energy density $\epsilon$ and sound speed $c_s$ could be also obtained systematically,
\begin{eqnarray}
\frac{dp(T)}{dT}&=& s(T),\\
\epsilon&=&-p+sT,\\
c_s^2&=&\frac{d \log T}{d \log s}=\frac{s}{T ds/dT}.
\end{eqnarray}

We have compared the results of the entropy density, the pressure density and sound velocity square with  the pure SU(3) lattice data from \cite{LAT-EOS-G} in \cite{Li:2014hja}, and it is found that these thermodynamical quantities obtained from the quenched dynamical holographic QCD model can agree with lattice calculation in pure gluon system. Here we only show the trace anomaly $\epsilon-3p$ with the pure SU(3) lattice data from \cite{LAT-EOS-G} in Fig.\ref{TraceA}, we can see that  the sharp peak of trace anomaly around the critical temperature $T_c$ has been reproduced successfully in this model. All these facts show that the dynamical model has encoded the correct IR physics.

\subsection{Jet quenching parameter $\hat{q}$}

Jet quenching measures the energy loss rate of an energetic parton going thorough the created hot dense medium. In \cite{Li:2014hja}, authors have had a detailed study on the temperature dependence behavior of $\hat{q}$ following the method in \cite{Liu:2006ug} (see also \cite{Gursoy:2009kk,Cai:2012eh,Zhang:2012jd}) in this model. $\hat{q}$ could release the signals of phase transition.

\begin{figure}[h]
\begin{center}
\epsfxsize=7.5 cm \epsfysize=6.5 cm \epsfbox{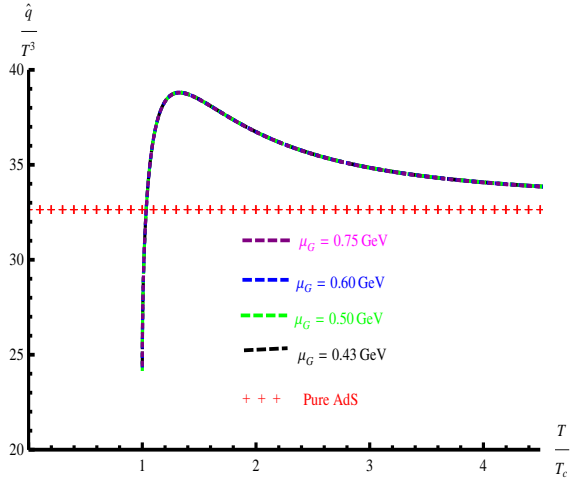}
\end{center}
\caption{$\hat{q/T^3}$ as a function of temperature $T$ with $G_5=1.25$ and $\lambda=6\pi$.} \label{qhatdT3}
\end{figure}

Following \cite{Liu:2006ug}, the jet quenching parameter is related to the
adjoint light like Wilson loop by the equation
\begin{eqnarray}
W^{Adj}[\mathcal {C}]\approx exp(-\frac{1}{4\sqrt{2}}\hat{q}L^{-}L^2).
\end{eqnarray}

The expectation value of Wilson loop is dual to the on-shell value of the string Nambu-Goto action with proper string configuration. The $\hat{q}$ can be obtained
\begin{eqnarray}\label{qhatfor-res}
\hat{q}=\frac{\sqrt{2}\sqrt{\lambda}}{\pi z_h^3 \int_0^{1}d\nu\sqrt{\frac{e^{-4A_s(\nu z_h)}}{z_h^4}\frac{1-f(\nu z_h)}{2}f(\nu z_h)}}.
\end{eqnarray}

Insert background solution Eqs.(\ref{quadratic-dilaton},\ref{solu-f},\ref{As-sol}) into the above equation (\ref{qhatfor-res}), one can get the temperature dependence of the jet quenching parameter numerically. In the temperature range $300 \sim 400 {\rm MeV}$,  the value of ${\hat q}$ is around $5\sim 10 {\rm GeV}^2/{\rm fm}$, which is in agreement with the lattice results in \cite{Panero:2013pla}. In  Fig.\ref{qhatdT3} we show $\hat{q}/T^3$ as a function of temperature, and it is seen that there is a peak with the height around 40 at around $T=1.1 T_c$. This behavior is quite different from the one obtained in pure AdS background, where $\hat{q}/T^3$ is a constant for all temperatures. This shows that the dynamical holographic QCD has encoded novel property about the deconfinement phase transition.

\section{Transport coefficients in a dynamical holographic QCD model}
\label{sec-tans-coeff}

We have already set up holographic model and we have seen that the jet quenching parameter over temperature cubic can reflect the phase transition. In the following, we would like to study various transport coefficients and check whether other transport coefficients can also characterize the deconfinement phase transition.

\subsection{Bulk viscosities}

There are many effective models, such as dissipative hydrodynamics which are used to understand the real-time evolution of the plasma. In these models, transport coefficients would encode the low frequency fluctuations. In this section, we focus on the temperature dependence behavior of the bulk viscosity. There are several approaches for calculating bulk viscosity in holographic framework \cite{Buchel:2009bh,Parnachev:2005hh,Benincasa:2005iv,Buchel:2005cv,Benincasa:2006ei,Mas:2007ng}. In this quenched dynamical holographic QCD model, it's convenient to follow the method \cite{Gubser:2008yx} and there is general derivation for graviton-dilaton system. Here we will briefly review how to extract the bulk viscosity in graviton-dilaton system and then we figure out the results for the model (\ref{quadratic-dilaton}, \ref{solu-f}, \ref{As-sol}).

In 4D field theory side, one possible way to extract the bulk viscosity is using the Kubo formula
\begin{eqnarray}
\zeta=\frac{1}{9}\underset{\omega\rightarrow0}{\text{lim}}\frac{1}{\omega}\text{Im}G^R_{xx,xx}(\omega),
\end{eqnarray}
where $G^R(\omega)$ is the retarded Green function of the stress energy tensor defined as
\begin{eqnarray}
G^R_{ij,kl}(\omega)=-i \int d^3xdt e^{i\omega t}\theta(t)\langle [T_{ij}(t,\overset{\rightarrow}{x}),T_{kl}(0,0)]\rangle,
\end{eqnarray}
with $T_{ij}$ the $i-j$ component of the stress tensor. According to the holographic dictionary, one can extract the Green function of the stress tensor in terms of metric perturbation like $g_{\mu\nu}\rightarrow g_{\mu\nu}+h_{\mu\nu}$ in the bulk. Here, in order to apply the Kubo formula, we choose the spatial component of momentum $\overset{\rightarrow}{q}=0$ and we also assume that $h_{\mu\nu}$ depends on $t, z$ only, i.e. $h_{\mu\nu}=h_{\mu\nu}(t,z)$. These metric perturbation components should contain $h_{xx},h_{yy},h_{zz}$. We assume that spatial rotation symmetry and $h_{xx}=h_{yy}=h_{zz}$. In principle, scalar modes like $h_{tz},h_{tt},h_{zz}$ and $\delta \Phi$ are all coupled with each other, which makes the calculation quite complicated. Fortunately, as pointed in \cite{Gubser:2008yx}, one can use the gauge symmetry to eliminate $h_{tz}$ and $\delta \Phi$ and only diagonal mode will survive finally. With following these logics, we can obtain bulk viscosity from retarded green function of scalar modes. The imaginal part of the retarded Green function $G_R$ is related to conserved flux
$\mathcal{F(\omega)}$
\begin{eqnarray}
\text{Im}G_{R}=-\frac{\mathcal{F(\omega)}}{4\pi G_5}.
\end{eqnarray}
Where the metric ansatz is
\begin{eqnarray}\label{metric-new}
ds^2=e^{2A}(-fdt^2+dx_idx^i)+\frac{e^{2B}}{f}dz^{2}
\end{eqnarray}
and coordinate definition $\phi(z)=z$, the conserved flux takes the form
\begin{eqnarray}
\mathcal{F}(\omega)=\frac{e^{4A-B}f}{4A^{'2}}|\text{Im} h_{xx}^{*}h_{xx}^{'}|.
\end{eqnarray}
Here, it should be noticed that these results are derived starting from the following Einstein frame action
\begin{eqnarray}\label{graviton-dilaton-E}
S_G=\frac{1}{16\pi G_5}\int d^5x\sqrt{g_E}\left(R_E-\frac{1}{2}\partial_m\phi\partial^m\phi-V(\phi)\right).
\end{eqnarray}
Therefore, before we use these formulae, we have to notice that there's a transformation from our convention $\Phi$ to $\phi$ by $\phi=\sqrt{\frac{8}{3}}\Phi=\sqrt{\frac{8}{3}} \mu_G^2 z^2$ and the metric in Eq.(\ref{metric-new}) is in Einstein frame.

An convenient property of the equation of motion of the perturbation is that the equation of motion for $h_{xx}$ component is simply decoupled from $h_{zz},h_{tt}$ and takes the following form
\begin{eqnarray}\label{hxx-eom}
h_{xx}^{''}=(-\frac{1}{3A^{'}}-4A^{'}+3B^{'}-\frac{f^{'}}{f})h_{xx}^{'}+(-\frac{e^{-2A+2B}}{f^{2}}\omega^2+\frac{f^{'}}{6f A^{'}}-\frac{f^{'}B^{'}}{f})h_{xx}.\nonumber\\
\end{eqnarray}

\begin{figure}[h]
\begin{center}
\epsfxsize=6.5 cm \epsfysize=6.5 cm \epsfbox{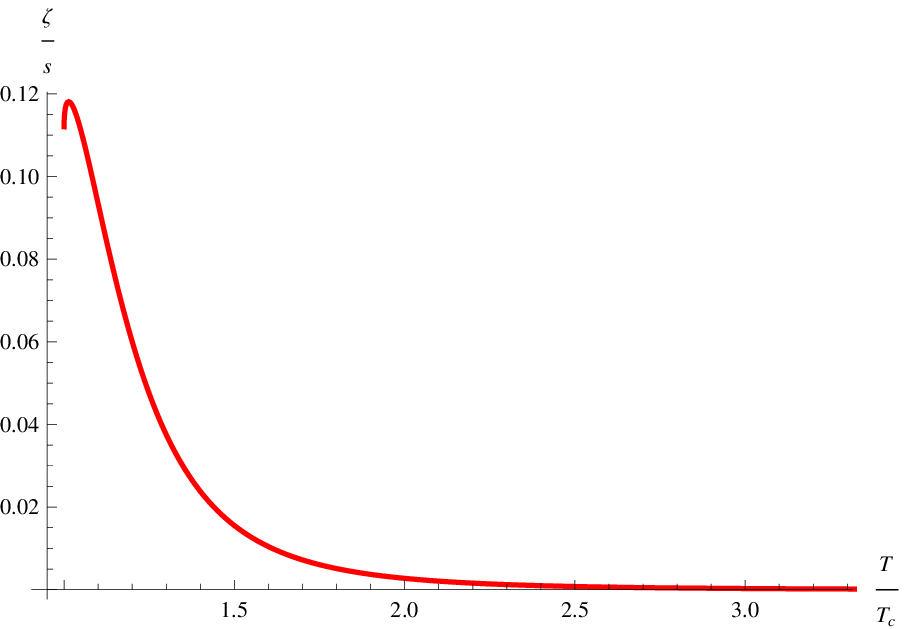} \hspace*{0.1cm}
\epsfxsize=6.5 cm \epsfysize=6.5 cm \epsfbox{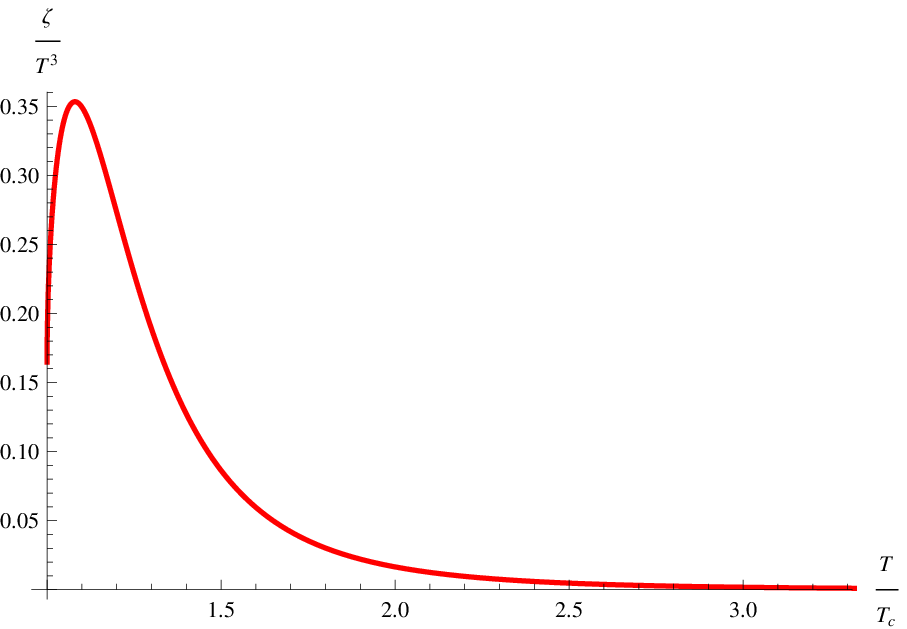} \vskip -0.05cm
\hskip 0.15 cm
\textbf{( a ) } \hskip 6.5 cm \textbf{( b )} \\
\end{center}
\caption[]{Bulk viscosity results when $\mu_G=0.75{\rm GeV}$ and $G_5=1.25$ } \label{zetas}
\end{figure}

These results are quite general within graviton-dilaton system. Therefore, it's very easy to extract the bulk viscosity in our model defined in Eqs.(\ref{quadratic-dilaton},\ref{solu-f},\ref{As-sol}) from the above formulae. In order to apply this analysis to our case, the only thing is to extract the $A,B$ factor in Eq.(\ref{metric-new}). Firstly, we take a new fifth coordinate as $z^{'}=Z(z)\equiv\sqrt{\frac{8}{3}}\mu_G^2z^2$, which satisfies the gauge choice $\phi(z^{'})=z^{'}$. We just rewrite down the new metric form in terms of the new coordinate $z^{'}$
\begin{eqnarray}
ds^2=e^{2A_s-\frac{4}{3}\Phi}(-fdt^2+dx_idx^i)+\frac{e^{2A_s}}{fZ^{'2}}dz^{'2}.
\end{eqnarray}
One can see $e^{2A}=e^{2A_s-\frac{4}{3}\Phi}$ and $e^{2B}=\frac{e^{2A_s-\frac{4}{3}\Phi}}{Z^{'2}}$. we insert these results into Eq.(\ref{hxx-eom})and solve the equation (\ref{hxx-eom}). The temperature behavior of the bulk viscosity is shown in Fig.\ref{zetas}(a) and (b). In Fig.\ref{zetas}(a), we have scaled the bulk viscosity in terms of the entropy density. In Fig.\ref{zetas}(b), we have scaled it with respect to $T^3$. In these two figures, there is a maximal point of bulk viscosity near the transition temperature. This kind of behavior is in agreement with the results in \cite{LAT-xis-Meyer,LAT-xis-KT}. More importantly, the near $T_c$ peak is similar to the behavior of trace anomaly results as shown in Fig.\ref{TraceA}, which shows that like jet quenching parameter, the near $T_c$ behavior of bulk viscosity also contains information of phase transition.

\subsection{Electric DC conductivity $\sigma_{0}$}

In this section, we would like to study DC conductivity $\sigma_0$ which may be helpful to understand the phase transition appeared in this model.
The DC conductivity $\sigma_0$ in 4D field theory is related to the retarded Green function $G^{R,EM}$ of electric-magnetic current $J^{EM}_\mu$ by $\sigma_{0}=-\underset{\omega\rightarrow0}{\text{lim}}\frac{1}{\omega}\text{Im} G_{xx}^{R,EM}$. In order to consider the current-current Green function, one has to introduce the $U(1)$ gauge field perturbation. As in \cite{DeWolfe:2011ts}, we will add the following probe action to the original action,

\begin{eqnarray}
S_F=-\frac{1}{16\pi G_5} \int d^5 x G(\Phi)\frac{F^2}{4},
\end{eqnarray}
Where we choose coupling form as $G(\Phi)=e^{-\Phi}$ in this paper. For simplifying analysis, we introduce a electric field perturbation $a=A_1$ with field strength $F_{\mu\nu}$. Following method given in \cite{DeWolfe:2011ts}, the equation of motion for $a$ in this model Eq.(\ref{metric-new}) is
\begin{eqnarray}
a^{''}+(2A^{'}-B^{'}+\frac{f^{'}}{f}+\frac{G^{'}\Phi^{'}}{G})a^{'}+\frac{e^{2B-2A}}{f^2}\omega^2 a=0.
\end{eqnarray}

With imposing $A,B$, we could extract the numerical results from
\begin{eqnarray}
\sigma_0=-\frac{1}{16\pi G_5}\underset{\omega\rightarrow0}\lim\frac{f G(\Phi)e^{2A-B}\text{Im}(a^{*} a^{'})}{\omega}.
\end{eqnarray}

\begin{figure}[h]
\begin{center}
\epsfxsize=7.5 cm \epsfysize=6.5 cm \epsfbox{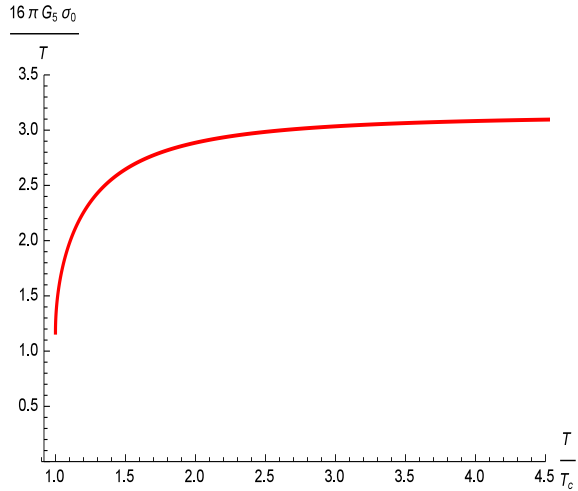}
\end{center}
\caption{DC conductivity $\sigma_0/T$ as a function of temperature $T$ with $G_5=1.25$ and $\mu_G=0.75GeV$.} \label{sigmadc}
\end{figure}

The $\sigma_{0}$ as function of temperature is shown in Fig.\ref{sigmadc}. Focusing on the near $T_c$ qualitative behavior, we could see that the dimensionless combination $\sigma_{el}/T$ decrease also very quickly approaching $T_c$ from above. This in some sense could be seen as a signal of phase transition.

\subsection{Shear viscosity with higher derivative corrections}

Shear viscosity is also one interesting transport coefficient in QGP. In this section, we focus on shear viscosity in the holographic dynamical model. It has been well studied in Einstein gravity background and it is universally proportional to the entropy density, i.e. $\eta/s=1/(4\pi)$ \cite{Policastro:2001yc,Buchel:2003tz}. To introduce temperature dependence of ratio between shear viscosity and entropy density, one has to introduce higher derivative gravity \cite{Cremonini:2012ny} corrections to the Einstein gravity. In order to obtain the temperature dependence of the ratio, we should also introduce the higher derivative gravity correction in this model and we will see what will happen in the model with higher derivative gravity correction finally. We introduce higher derivative correction with following form

\begin{eqnarray}
S=\frac{1}{16\pi G_5}\int d^5x \sqrt{-g}\big(&R&-\frac{4}{3}\partial_\mu\Phi\partial^\mu \Phi-V(\Phi)\nonumber\\
&+&\beta e^{\sqrt{2/3}\gamma\Phi}R_{\mu\nu\lambda\rho}R^{\mu\nu\lambda\rho}\big).
\end{eqnarray}
Here the extra $\sqrt{2/3}$ factor in the coupling of $R_{\mu\nu\lambda\rho}R^{\mu\nu\lambda\rho}$ is to keep our parameter$\gamma$ comparable to the one in \cite{Cremonini:2012ny}. The $\beta$ is small dimensionless parameter. In order to calculate shear viscosity, one must introduce the off diagonal metric perturbation $h_{xy}$ to obtain two point Green function of the energy momentum tensor. The shear viscosity could be calculated by Kubo formula
\begin{eqnarray}
\eta=-\underset{\omega\rightarrow0}{\text{lim}}\frac{1}{\omega}\text{Im}G^R_{xy,xy}(\omega).
\end{eqnarray}

\begin{figure}[h]
\begin{center}
\epsfxsize=7.5 cm \epsfysize=5.5 cm \epsfbox{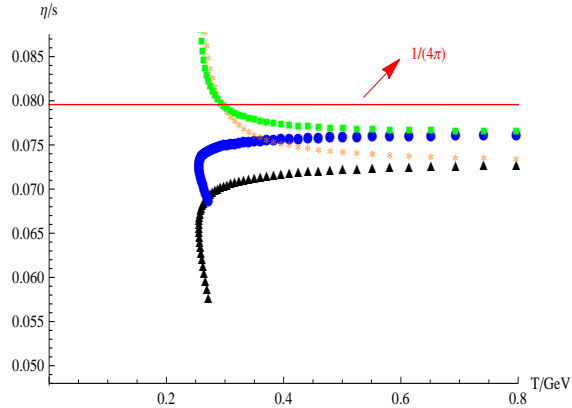}
\end{center}
\caption[]{$\eta/s$ results for different parameter values. The green dots stand for $\beta=0.005,\gamma=\sqrt{2/3}$, the orange dots stand for $\beta=0.01,\gamma=\sqrt{2/3}$, the blue dots stand for $\beta=0.005,\gamma=-\sqrt{2/3}$ and the black dots stand for $\beta=0.01,\gamma=-\sqrt{2/3}$. } \label{etads-para}
\end{figure}

In principle, after introducing the higher derivative corrections, the background metric would be deformed in $O(\beta)$ level. Fortunately, as pointed out in \cite{Cremonini:2012ny},  $\eta/s$ would only depend on the $O(1)$ metric background. In this sense, we would not consider the ratio with corrections from the deformation of the background metric up to $O(\beta)$. Following the arguments \cite{Cremonini:2012ny}, the shear viscosity over entropy density ratio results read up to $O(\beta)$
\begin{eqnarray}
\eta/s=\frac{1}{4\pi}\left(1-\frac{\beta}{c_0}e^{\sqrt{2/3}\gamma\Phi_h}(1-\sqrt{2/3}\gamma z_h\Phi^{'}(z_h))\right)
\end{eqnarray}
with $c_0=-z_h^5\partial_z\left((1-z^2/z_h^2)^2e^{2A_s-\frac{4}{3}\Phi}/(8f(z)z^2)\right)|_{z=z_h}$. The numerical results about the ratio with different parameter values are shown in Fig.\ref{etads-para}. Fig.\ref{etads-para} shows that the qualitative behavior of $\eta/s$ is sensitive to changes of parameter. The ratio highly depends on the temperature as we expected. At the large temperature region, $\eta/s$ would tend to a constant, the value of which would depend on $\beta, \gamma$. Near $T_c$, if $\beta\neq0$, $\eta/s$ would rise quickly if $\gamma>0$ from above, while for $\gamma<0$, the curves bent down.

\begin{figure}[h]
\begin{center}
\epsfxsize=6.5 cm \epsfysize=4.5 cm \epsfbox{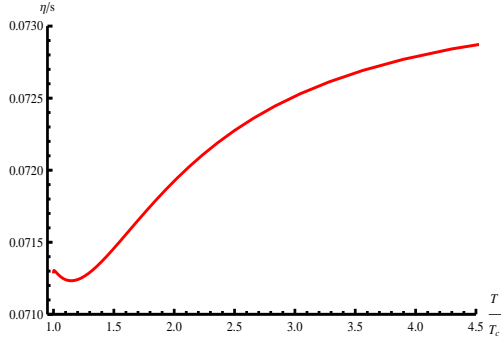}
\end{center}
\caption[]{$\eta/s$ results when $\beta=0.01,\gamma=-\sqrt{8/3}$.} \label{etads-valley}
\end{figure}

When $\beta=0.01,\gamma=-\sqrt{8/3}$, we show the ratio of $\eta/s$ as a function of $T/T_c$ in Fig.\ref{etads-valley}. It is seen that the ratio $\eta/s$  shows a valley around $T=1.1T_c$. The location of valley is almost same as the location of the peak position in $\hat{q}/T^3$
and the peak of the trace anomaly as well as the peak of the bulk viscosity.

\begin{figure}[h]
\begin{center}
\epsfxsize=6.5 cm \epsfysize=4.5 cm \epsfbox{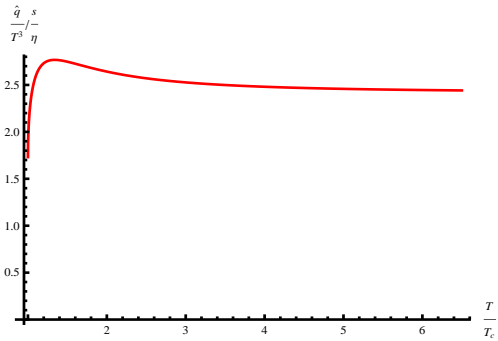}
\end{center}
\caption[]{$\hat{q}/T^3*\eta/s$ results when $\beta=0.01,\gamma=-\sqrt{8/3}$.} \label{qhatds-d-sdeta}
\end{figure}

In Fig.\ref{qhatds-d-sdeta}, we show the configuration of $\hat{q}/T^3*\eta/s$ with respect to temperature in this model. When $T$ becomes large, $\hat{q}/T^3*\eta/s$ reaches a constant 2.5, which is 2 times larger than the perturbative result given in \cite{Majumder:2007zh}. Near $T_c$, though the value is not constant, by comparing Fig.\ref{etads-valley} and the results of $\hat{q}/T^3$ in Fig.\ref{qhatdT3}\footnote{Since $\hat{q}/T^3$ is dimensionless and much larger than $1$ and $\beta$ is of order $0.01$, we would expect that the O($\beta$) correction would not change the result of $\hat{q}/T^3$ much and use the results calculated in previous section.}. It could be easily seen that when $\hat{q}/T^3$ increase that $\eta/s$ decrease, which could be seen as the strong coupling extension of the conclusion in \cite{Majumder:2007zh}. This result can be check by lattice calculation.

\section{Conclusion and discussion}
\label{sec-sum}

In this paper, based on the dynamical holographic QCD model constructed in \cite{Li:2014hja,Li:2012ay,Li:2013oda}, we investigate the temperature dependent behavior of the transport coefficients, including the bulk/shear viscosities, the electric conductivity and the jet quenching parameter. Based on the previous constraints from the equation of states and meson spectra, our model can produce a reasonable results on the transport coefficients. Furthermore, we show that in the near $T_c$ region around $T=1.1T_c$, the dimensionless combinations of transport coefficients like $\eta/s,\zeta/s,\hat{q}/T^3,\sigma_{el}/T$ would change sharply, which reveals the information of phase transition.

Besides, near $T_c$ and around $T=1.1T_c$, we find that the location of the peak in $\hat{q}/T^3$ and the local minimal position of $\eta/s$ are coincident with each other. Though the relation between $\hat{q}/T^3$ and $\eta/s$ is no longer simple as in weakly coupled region, the inverse correlating relation could also be considered as the strong coupling extension of the conclusion on the strong relation of the two in weakly coupled regime in \cite{Majumder:2007zh}.

\vskip 0.5cm
{\bf Acknowledgement}
\vskip 0.2cm
 We thank Li Li,Yun-long Zhang for useful discussions. SH is supported by JSPS postdoctoral fellowship for foreign researchers and by the National Natural Science Foundation of China (No.11305235). SH is grateful to Tadashi Takayanagi for the support. MH is supported by the NSFC under Grant Nos. 11275213, 11261130311(CRC 110 by DFG and NSFC), CAS key project KJCX2-EW-N01, and Youth Innovation Promotion Association of CAS.

\end{document}